\documentclass[12pt]{emulateapj} 
\usepackage{amsmath}
\usepackage{amssymb}
\usepackage{amstext}
\usepackage{graphicx}
\usepackage{epstopdf}
\usepackage{epsfig}
\usepackage{url}
\usepackage{color}
\definecolor{myColor}{rgb}{0.9,0.9,0.9}  
\begin{document}
\renewcommand\bottomfraction{.9}
\shorttitle{Characterizing Water-rich Super-Earths} 
\title{Optimal Measures for Characterizing Water-rich Super-Earths} 
\author{Nikku Madhusudhan\altaffilmark{1,2} \& Seth Redfield\altaffilmark{3}}
\altaffiltext{1}{Institute of Astronomy, University of Cambridge, Cambridge CB3 0HA, UK {\tt nmadhu@ast.cam.ac.uk}}
\altaffiltext{2}{Formerly at the Department of Physics and Department of Astronomy, Yale University, New Haven, CT 06511}
\altaffiltext{3}{Astronomy Department, Van Vleck Observatory, Wesleyan University, Middletown, CT 06459, USA}

\begin{abstract}
The detection and atmospheric characterization of super-Earths is one of the major frontiers of exoplanetary science. Currently, extensive efforts are underway to detect molecules, particularly H$_2$O, in super-Earth atmospheres. In the present work, we develop a systematic set of strategies to identify and observe potentially H$_2$O-rich super-Earths that provide the best prospects for characterizing their atmospheres using existing instruments. Firstly, we provide analytic prescriptions and discuss factors that need to be taken into account while planning and interpreting observations of super-Earth radii and spectra. We discuss how observations in different spectral bandpasses constrain different atmospheric properties of a super-Earth, including radius and temperature of the planetary surface as well as the mean molecular mass, the chemical composition and thermal profile of the atmosphere. In particular, we caution that radii measured in certain bandpasses can induce biases in the interpretation of the interior compositions. Secondly, we investigate the detectability of H$_2$O-rich super-Earth atmospheres using the HST WFC3 spectrograph as a function of the planetary properties and stellar brightness. We find that highly irradiated super-Earths orbiting bright stars, such as 55 Cancri e, present better candidates for atmospheric characterization compared to cooler planets such as GJ~1214b even if the latter orbit lower-mass stars. Besides being better candidates for both transmission and emission spectroscopy, hotter planets offer higher likelihood of cloud-free atmospheres which aid tremendously in the observation and interpretation of spectra. Finally, we present case studies of two super-Earths, GJ~1214b and 55 Cancri e, using available data and models of their interiors and atmospheres.
\end{abstract} 

\keywords{planetary systems --- planets and satellites: general --- planets and satellites: individual (GJ~1214~b, 55~Cancri~e)}

\section{Introduction} 
\label{sec:intro}

The holy grail of exoplanetary science is ultimately the detection and atmospheric characterization of an Earth analogue. Recent observational surveys have already detected transiting exoplanets with terrestrial-like masses and/or radii (e.g. Leger et al. 2009; Batalha et al. 2011; Barclay et al. 2013; Borucki et al. 2013), and a wide range of equilibrium temperatures, including some in the habitable zones of their host stars (e.g. Borucki et al. 2013, Quintana et al. 2014). To date, masses and radii have both been measured for about 20 transiting super-Earths, defined as planets with masses between 1 and 10 Earth Masses (Valencia et al. 2006; Seager et al. 2007). Furthermore, exoplanet occurrence rates derived from surveys are revealing that sub-Neptune size planets are the most numerous class of planets in the solar neighborhood (Howard et al. 2012; Fressin et al. 2013). Currently, characterizing the atmospheres of such low-mass planets is one of the most active frontiers of exoplanetary science. 

Atmospheric characterization of super-Earths with current and upcoming facilities requires a focused assessment of objectives. Observational surveys increasingly desire to find super-Earths in the habitable zones of their host stars, and, if the planets happen to be transiting, to characterize their atmospheres. Our notions of habitability are commonly based on equilibrium temperatures (T$_{\rm eq}$) where liquid water can sustain, assuming that H$_2$O is indeed abundant in the planets in the first place (Kasting 1993; Selsis 2007; Abe et al. 2011; Kaltenegger et al. 2013; Kopparapu et al. 2013a,b). However, testing this assumption by observationally detecting H$_2$O in the atmosphere of a habitable super-Earth is beyond the reach of current observational facilities, and would be challenging even with larger forthcoming facilities within this decade (Kaltenegger \& Traub 2009; Belu et al. 2011,2013; Hedelt 2013; Snellen et al. 2013). Therefore, currently there is no plausible means for directly assessing the atmospheric chemical compositions, and hence the true habitability, of habitable-zone super-Earths and terrestrial analogs. 

A more achievable goal at the present time is to answer the more basic question of what is the frequency of H$_2$O-rich super-Earths irrespective of whether their temperatures are habitable or not. Such a question opens up the sample space to short-period transiting super-Earths whose atmospheres can potentially be characterized with existing facilities. While the short periods increase the probability and frequency of transits, the higher atmospheric temperatures make them more favorable for detecting H$_2$O in their atmospheric spectra. The outlook for characterization of such super-Earths is promising given that upcoming surveys from space, such as TESS (Ricker et al. 2014), CHEOPS (Broeg et al. 2013), and PLATO (Rauer et al. 2013), and on ground (e.g. Snellen et al. 2012; Gillon et al. 2013a) are expected to find large numbers of short-period super-Earths orbiting bright and low-mass stars. 

In the present work, we develop a framework for identification and characterization of H$_2$O-rich super-Earths with existing observational facilities. We present analytic prescriptions and theoretical results that are useful for planning and interpretation of super-Earth observations. We demonstrate how upper-limits on the atmospheric mean molecular mass can be derived for certain super-Earths based only on their masses and radii, and that only radii measured in certain bandpasses (`opacity windows') can be used to derive unbiased constraints on their interior compositions. We also explore the dependence of super-Earth spectra on chemical composition and temperature. And, finally we investigate the detectability of H$_2$O-rich super-Earth atmospheres with the Hubble Space Telescope ({\it HST}) WFC3 Spectrograph, and derive sensitivity estimates for super-Earths orbiting two broad stellar prototypes, a G dwarf and an M dwarf, over a wide range of planetary equilibrium temperatures (T$_{\rm eq}$) and stellar brightnesses. 

\section{Constraints from Mass and Radius}
\label{sec:interior}

\begin{figure}[t]
\centering
\includegraphics[width = 0.5\textwidth]{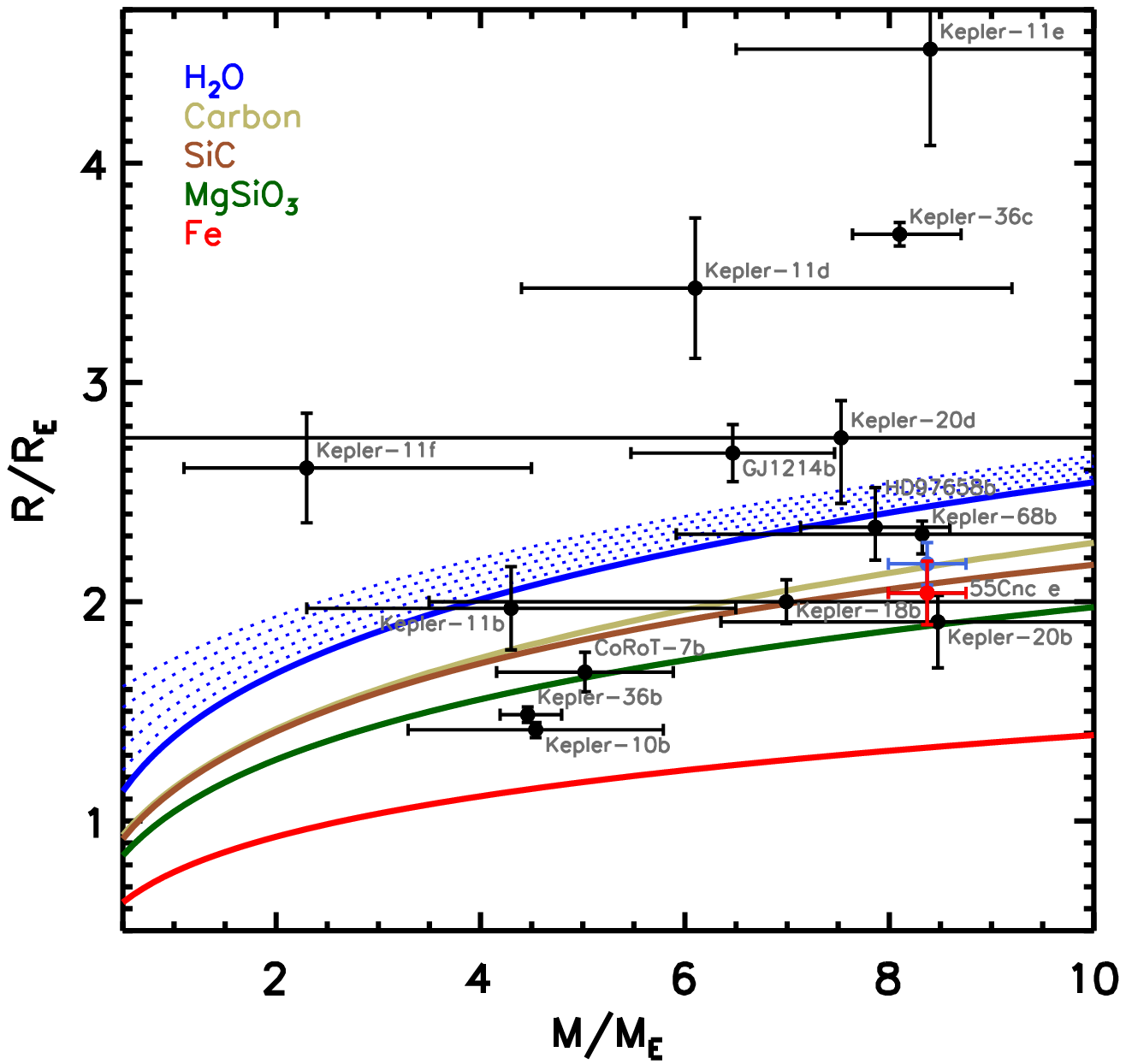}
\caption{Masses and radii of super-Earths and theoretical mass-radius relations. The colored solid curves show mass-radius relations predicted by internal structure models of planets with different uniform compositions shown in the legend in the same order as the curves, i.e. Fe (bottom-most curve) to H$_2$O (top-most curve). The model curves are from Madhusudhan et al. (2012). The blue dotted curves show maximum-radius curves (see section~\ref{sec:m-r_atmos}) for pure-H$_2$O planets with H$_2$O-rich atmospheres of different temperatures (500 K, 1000 K, 1500 K, 2000 K, and 2500 K) as observed in absorption bands of H$_2$O in transit; curves with larger radii correspond to higher temperatures. The black circles with error bars show measured masses and radii of known transiting super Earths ($M_p=1-10$ M$_\oplus$), adopted from the exoplanet orbit database (Wright et al. 2011). Two values of radii are shown for 55~Cancri~e. The red data point shows the radius measured in the visible, and the blue data point shows a ÒgrayÓ radius obtained by combining visible and infrared measurements (Winn et al. 2011; Gillon et al. 2012). 
}
\label{fig:m-r}
\end{figure}

The observed mass and radius ($M_p$, $R_p$) of a super-Earth can be used to place nominal constraints on its interior and atmospheric composition using internal structure models. Figure~\ref{fig:m-r} shows mass-radius relations for homogeneous planets of various compositions, along with the masses and radii of several  transiting super-Earths. The internal structure model and mass-radius curves are described in Madhusudhan et al. (2012). The compositions shown in Figure~\ref{fig:m-r} include the most common minerals typically invoked for super-Earth interiors, namely, Fe, silicates, and H$_2$O (Valencia et al. 2006; Seager et al. 2007; Sotin et al. 2007; Fortney et al. 2007).  Also shown are curves for SiC and C which, though rarely expected, can nevertheless be abundant in C-rich environments (Madhusudhan et al. 2012). Given the mass of a super-Earth, its radius can be explained by an often degenerate set of solutions comprising of various proportions of the different minerals listed above (e.g. Rogers \& Seager 2009; Valencia et al. 2010; Wagner et al. 2012; Gong \& Zhou 2012; Madhusudhan et al. 2012). 

Despite the degeneracy in solutions, the likelihood of a H$_2$O-rich atmosphere in a super-Earth can still be assessed from $M_p$ and $R_p$. We define three super-Earth types in this regard (SE1, SE2, SE3). For type SE1, planets with $M_p$ and $R_p$ lying between the iron and silicate curves, a wide range of compositions are possible, including water-rich and water-poor conditions. On the other hand, for type SE2 planets with $M_p$ and $R_p$ lying between the silicate and H$_2$O curves, a volatile-rich envelope (e.g. made of H/He, H$_2$O, etc.) is required to explain the density, unless, in rare cases, where a carbon-rich composition can be invoked based on the stellar abundances (e.g. Madhusudhan et al. 2012; Moriarty et al. 2014). For typical O-rich host stars, therefore,  type SE2 planets are good candidates for hosting H$_2$O-rich envelopes and atmospheres. And, finally, type SE3  super-Earths with $M_p$ and $R_p$ lying above the H$_2$O curve, necessarily require an envelope composition lighter than H$_2$O, most likely a H/He envelope and an atmosphere (e.g. Seager et al. 2007). Using these attributes, the atmospheric mean molecular mass ($\mu$) of super-Earths of different types can be constrained in different ways as discussed below. 

\subsection{Upper-limit on $\mu$ of {\rm SE3}-type Atmospheres}  

In SE3-type planets, which are expected to host volatile-rich envelopes, nominal constraints can be placed on the $\mu$ of the atmosphere from the mass and a monochromatic radius. For such super-Earths, a minimum thickness of the atmosphere can be defined as the difference between the observed radius ($R_p$) and the radius of a 100\% water planet ($R_{\rm H_2O}$) with the observed mass ($M_{\rm p}$) (Kipping et al. 2013). Expressing the atmospheric thickness ($H$) in a molecular spectral band in terms of the scale height ($H_{\rm sc}$) of the atmosphere gives 

\begin{equation}
R_{\rm p} - R_{\rm H_2O}(M_{\rm p}) = N_{\rm sc}H_{\rm sc} = N_{\rm sc}k_{\rm b}T/ \mu g, 
\label{eq:rp1}
\end{equation}
where, $N_{sc}$ is the number of scale heights of the atmosphere contributing to the effective radius of the planet at the observed wavelength, and is typically of the order 5-10 (see section \ref{sec:transmission}). Here, $\mu$ is the mean molecular mass of the atmosphere, $g$ is the acceleration due to gravity, $k_{\rm b}$ is the Boltzmann constant, and $T$ is a characteristic temperature of the atmosphere at the day-night terminator of the planet. Equation~(\ref{eq:rp1}) can be used to derive a nominal upper-limit on the atmospheric $\mu$ as 
\begin{equation}
\mu \lesssim \mu_{\rm max} = \frac{10~k_{\rm b}T_{\rm eq}}{g_0(M_{\rm p})[R_{\rm p} - R_{\rm H_2O}(M_{\rm p})]}. 
\label{eq:rp2}
\end{equation}

Here, a characteristic $T$ of $T_{\rm eq}$ and a maximal $N_{sc}$ of 10 are assumed, where $T_{\rm eq}$ is the equilibrium temperature of the planet assuming full redistribution (see e.g. Madhusudhan 2012), and $g_0(M_{\rm p})$ is the acceleration due to gravity of a pure-H$_2$O planet with a mass $M_{\rm p}$ and is given by $G M_{\rm p}/R_{\rm H_2O}^2$, where $G$ is the gravitational constant.  

\subsection{Upper-limit on $\mu$ of {\rm SE1} and {\rm SE2}-type planets} 

For planets in the SE1 and SE2 types, hardly any constraints can be placed on the atmospheric $\mu$ by a monochromatic radius measurement, as their masses and radii can be explained by various interior compositions, as discussed above. For planets in these classes, multi-wavelength observations of the radius (i.e. transmission spectra) are required to discern the presence of an atmosphere. The vertical extent of the atmosphere ($N_{sc} H_{sc}$ in Eq~\ref{eq:rp1}) for such planets is derived from the difference between radius measured in a molecular band ($R_{\rm p,molec}$) and that measured in a spectral band with no strong opacity ($R_{\rm p,0}$), i.e. an `opacity window' (see section \ref{sec:atmos_thickness}). The constraint on $\mu$ is then given by 
\begin{equation}
\mu \lesssim \mu_{\rm max} = \frac{10~k_{\rm b}T_{\rm eq}}{g_0(M_{\rm p},R_{\rm p,0})[R_{\rm p,molec} - R_{\rm p,0}]},  
\label{eq:rp2_1}
\end{equation}
where, $g_0(M_{\rm p},R_{\rm p,0})$ = $G M_{\rm p}/R_{\rm p,0}^2$. This approach which is applicable to all super-Earth types, is discussed in more detail section~\ref{sec:spectra}. 

\subsection{Optimal Mass-Radius Space for H$_2$O detectability}
\label{sec:m-r_atmos}
Given the mass and temperature of a super-Earth, Eq. (\ref{eq:rp2}) can be used to define an upper-limit on the observable radius in a H$_2$O absorption band for super-Earths with H$_2$O-rich atmospheres. The maximum possible radius of a H$_2$O-rich super-Earth of a given mass can be approximated by the sum of the radius of a pure-H$_2$O planet of the same mass and the maximum possible atmospheric thickness for a given planetary temperature:   

\begin{equation}
R_p \lesssim R_{\rm p,max}(M_{\rm p}) = R_{\rm H_2O}(M_{\rm p}) + 10~k_{\rm b}T_{\rm eq}/ 18ug_0.  
\label{eq:rp3}
\end{equation}

\noindent
Here, a pure H$_2$O atmosphere is assumed with a $\mu$ of 18 amu; $u$ is the atomic mass unit (amu). 

Therefore, given the $M_{\rm p}$ and $T_{\rm eq}$ of a super-Earth, its radius should lie below $R_{\rm p,max}$ for it to host a H$_2$O-rich atmosphere. Fig.~\ref{fig:m-r} shows curves of $R_{\rm p,max}$ for H$_2$O-rich planets for a wide range of temperatures ($T_{\rm eq} = 500 - 2500 $K) encompassing those of currently known super-Earths. These curves define the limits of $M_p - R_p$ space for detecting H$_2$O-rich super-Earths; i.e. planets with $R_p$ above a curve with the corresponding $T_{\rm eq}$ are unlikely to host H$_2$O-rich atmospheres. As an example, the super-Earth GJ~1214b with $T_{\rm eq} \sim 550$ K is less likely to host a H$_2$O-rich atmosphere (see section~\ref{sec:gj1214_interior} for detailed discussion).  Additionally, as shown in Fig.~\ref{fig:m-r}, the available $M_p - R_p$ space for detecting H$_2$O-rich super-Earths is larger for hotter planets. 

\section{Constraints from Atmospheric Spectra}
\label{sec:spectra}

In this section, we attempt to answer the question of which super-Earths are best targets for atmospheric characterization. Our goal here is to identify factors which influence the observable signal of a H$_2$O-rich atmosphere, so as to aid in selecting optimal super-Earths for follow-up observations and detailed atmospheric characterization. In order to model the spectra, we use the 1-D approach developed in Madhusudhan \& Seager (2009) which allows parametric prescriptions for the compositions and temperature structures, and is applicable over a wide range of temperatures and compositions (e.g. Madhusudhan \& Seager 2009; Madhusudhan 2012). 

\subsection{Transmission Spectra}
\label{sec:transmission}

\begin{figure}[t]
\centering
\includegraphics[width = 0.5\textwidth]{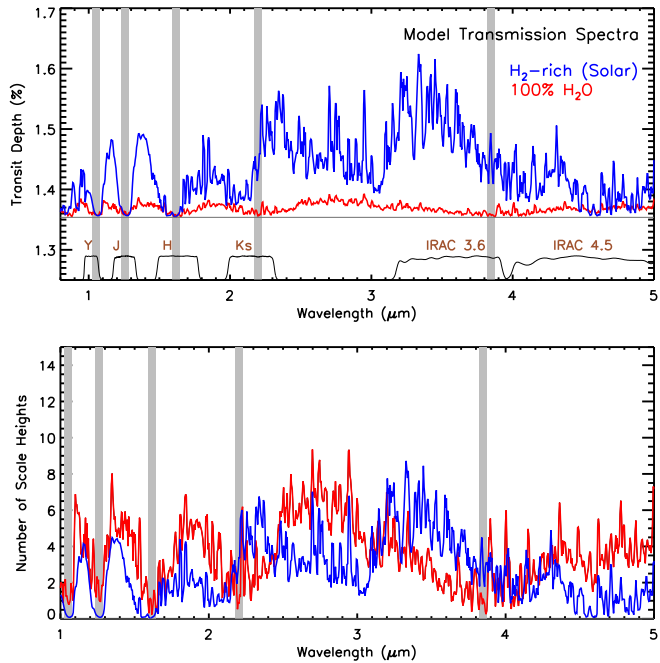}
\caption{Model transmission spectra with system parameters of GJ1214b. {\it Top Panel:} The blue and red curves show model transmission spectra for an atmosphere with H$_2$-rich and H$_2$O-rich composition. For the H$_2$-rich case, a solar abundance composition in thermochemical equilibrium was assumed. The horizontal gray line shows the transit depth corresponding to the bulk radius of the planet, i.e. the `surface', which can be observed in the opacity windows (see section~\ref{sec:surface_radius}). The gray vertical bands show opacity windows for a H$_2$O-rich atmosphere in the near-infrared, shown in 0.05$\mu$m-wide bands centered at 1.05, 1.26, 1.62, 2.20, and 3.85 $\mu$m. The black curves at the bottom show commonly used photometric bandpasses: $Y$, $J$, $H$, and $Ks$, which are accessible from ground, and {\it Spitzer} IRAC bands at 3.6 and 4.5 $\mu$m. {\it Bottom Panel:} The blue and red curves show the same spectra as in the top panel but in units of number of scale heights ($N_{sc,\lambda}$; see Eq.~\ref{eq:atmos1}) in the atmosphere above the `surface' i.e. the gray line in the top panel.}
\label{fig:gj1214_1}
\end{figure}

\begin{figure}[t]
\centering
\includegraphics[width = 0.5\textwidth]{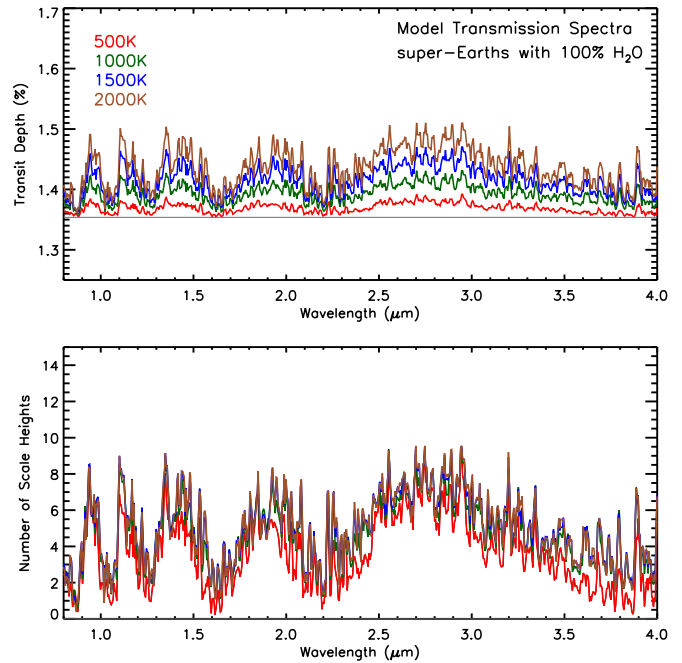}
\caption{Model transmission spectra of H$_2$O-rich super-Earths as a function of atmospheric temperature. {\it Top Panel:} All the spectra assume a H$_2$O-rich composition and isothermal temperature profiles with the specified temperatures. {\it Bottom Panel:} Spectra in units of $N_{sc,\lambda}$ corresponding to spectra in top panel (see caption of Fig.~\ref{fig:gj1214_1} for description). The models show that spectral features are enhanced with increasing temperature due to the increasing scale height (Eq.~\ref{eq:atmos1}), but the number of scale heights probed by the spectra is relatively independent of temperature. 
}
\label{fig:gj1214_2}
\end{figure}

A transmission spectrum, observed when the planet is in transit, probes the atmosphere near the day-night terminator of the planet. Several recent studies have investigated methods to use transmission spectra of super-Earths to constrain their various atmospheric properties, including mean-molecular masses, temperature profiles, and the presence of scatterers (e.g. Miller-Ricci et al. 2009; Benneke \& Seager 2012,2013; Howe \& Burrows 2012). Figures~\ref{fig:gj1214_1} and \ref{fig:gj1214_2} show model transmission spectra with the bulk parameters (planet radius and gravity, and stellar radius) of the GJ~1214b system, but exploring different chemical compositions and temperatures. For ease of illustration, in these models we assume isothermal temperature profiles, also because infrared transmission spectra are not strongly sensitive to the temperature profile at the terminator (Miller-Ricci \& Fortney 2010; Howe \& Burrows 2012). The H$_2$O-rich models comprise of 100\% H$_2$O, and the H$_2$-rich atmospheres comprise of solar abundance composition (Madhusudhan \& Seager 2011), i.e. H$_2$ and He constitute $\sim$99.9\% of the composition by volume but contribute minimal spectral features, and the rest in molecules such as H$_2$O, CH$_4$, CO, and CO$_2$ which contribute the prominent spectral features. 

\subsubsection{`Surface radius' of a Super-Earth}
\label{sec:surface_radius}
 Knowing the surface radius of a super-Earth is important to constrain its interior composition. Transit depths, or equivalently radii, of super-Earths have been reported in multiple bandpasses in the visible and infrared wavelengths. Radius measurements of a super-Earth at different wavelengths cannot be combined to improve upon the measurement uncertainties, because different spectral bandpasses encode information from different depths of the planetary atmosphere. On the other hand, radii of super-Earths are routinely used to constrain their interior compositions, irrespective of the observed spectral bandpass (e.g. Winn et al. 2011; Demory et al. 2011; Gillon et al. 2012). Such an exercise assumes that the radius used represents the bulk radius of the planet, without any contribution from an overlying atmosphere. Therefore, the resulting constraints on the interior composition can be erroneous if the adopted radius was measured at wavelengths where significant absorption from an atmosphere is possible. Thus, it is important to identify spectral bandpasses in which the radii measured represent the bulk `surface' radius ($R_{ps}$) of the planet and those in which the radii may include significant contribution from the atmosphere. Here, `surface' is nominally defined as the altitude where $\tau \sim 1$ when observed at wavelengths with minimal atmospheric opacity. In `opacity windows', where atmospheric molecular line absorption is minimal, the $\tau \sim 1$ surface may imply (a) a physical solid/liquid surface, (b) the deeper, high pressure, regions of a gaseous atmosphere where collision-induced absorption (CIA) may contribute significant opacity, or (c) the presence of a cloud deck aloft in the atmosphere. In the presence of CIA or cloud opacity, $R_{ps}$ represents only an upper-limit on the bulk radius of the planet. 

We identify spectral ranges of several opacity windows in Fig.~\ref{fig:gj1214_1} which provide ideal bandpasses at which to measure $R_{ps}$ of super-Earths (e.g. narrow bands at 1.05 $\mu$m, 1.26 $\mu$m, and 1.62 $\mu$m). As shown in the figure, all the spectra approach $R_{ps}$ of the planet at these wavelengths where there is minimal H$_2$O absorption and hence the starlight passes largely unimpeded through the planetary atmosphere. These bandpasses also coincide with the bands in which telluric H$_2$O contamination is minimal. As such, the conventional near-IR bandpasses of  $Y$(1.0 $\mu$m), $J$ (1.2 $\mu$m), $H$(1.6 $\mu$m), and $K$ (2.1 $\mu$m), which are accessible with ground-based facilities and partly with the {\it HST} WFC3 spectrograph, provide good bandpasses to measure surface radii of H$_2$O-rich super-Earths which can be used to constrain their bulk compositions. Ideally, however, narrower bands identified in Fig.~\ref{fig:gj1214_1} would provide the best estimates of $R_{ps}$. 
  
\subsubsection{Atmospheric Thickness and Transit Depth}
\label{sec:atmos_thickness}

The wavelength-dependent super-Earth radius ($R_{p\lambda}$) measured outside the opacity windows can include contributions from molecular opacity in the planet's atmosphere  and, hence, can be larger than the surface radius. The $\lambda$-dependent thickness of the atmosphere ($H_\lambda$), as alluded to in Eq (\ref{eq:rp1}), is given by 

\begin{equation}
H_\lambda = R_{p\lambda} - R_{ps} =  N_{sc, \lambda}k_{\rm b}T_{\rm eq}/ \mu g. 
\label{eq:atmos1}
\end{equation}

For a transiting super-Earth, the transit depth at primary eclipse is given by 
\begin{equation}
\delta_\lambda = R_{p\lambda}^2/R_{s\lambda}^2 = (R_{ps}  + H_\lambda)^2/R_{s\lambda}^2, 
\label{eq:atmos2}
\end{equation}
where, $R_{s\lambda}$ is the stellar radius. Therefore, the $\lambda$-dependent contribution of the atmosphere to the transit depth is given by 

\begin{equation}
\Delta_\lambda \sim 2 H_\lambda R_{ps} / R_{s\lambda}^2 = 2N_{sc, \lambda}k_{\rm b}T_{\rm eq}R_{ps}/ (\mu g R_{s\lambda}^2).  
\label{eq:atmos3}
\end{equation}
This quantity, $\Delta_\lambda$, constitutes the `signal' when planning observations of a super-Earth atmosphere. For a transiting super-Earth, $T_{\rm eq}$, $R_{ps}$, and $R_{s}$ are expected to be known from the system parameters. On the other hand, $N_{sc, \lambda}$ and $\mu$ which depend on the chemical composition of the planetary atmosphere, are not known a priori and, hence, need to be estimated based on theoretical models. For an assumed molecular composition, $\mu$ is known, e.g. 2.37 for a solar-composition (dominated by H$_2$ and He) atmosphere, 18 for 100\% H$_2$O, 44 for 100\% CO$_2$, etc. 

We use model spectra to estimate $N_{sc, \lambda}$ for a representative range of atmospheric compositions and temperatures. The lower panel of Fig.~\ref{fig:gj1214_1} shows $N_{sc, \lambda}$ for solar as well as H$_2$O-rich atmospheric compositions for model transmission spectra of super-Earth GJ~1214b. For both models, in the opacity windows $N_{sc, \lambda}$ is close to zero, leading to $H_\lambda \sim 0$ and $\Delta_\lambda \sim 0$, and hence $R_{p\lambda} = R_{ps}$. On the other hand, at wavelengths corresponding to the molecular absorption features, $N_{sc, \lambda}$ is non-zero and can be as high as 8 at the center of the absorption bands, depending on the strength of the feature. Therefore, in estimating transit depths of super-Earth atmospheres using Eq (\ref{eq:atmos1}) -- (\ref{eq:atmos3}), the following values of $N_{sc, \lambda}$ are recommended for a given chemical composition: 
\begin{eqnarray}
N_{sc, \lambda} \sim 0 ~~\textrm{(in opacity windows)} \nonumber \\
N_{sc, \lambda} \sim 5-8 ~~\textrm{(in molecular bands)}
\label{eq:Nsc_lambda}
\end{eqnarray}
These values of $N_{sc, \lambda}$ hold generally true irrespective of the atmospheric temperatures, as discussed in the following section~\ref{sec:atmos_temp} and in Fig.~\ref{fig:gj1214_2}. 

\subsubsection{Effect of Temperature}
\label{sec:atmos_temp}

For a super-Earth with a given radius and composition, hotter atmospheres are more conducive to atmospheric observations and characterization. As shown in Fig.~\ref{fig:gj1214_2}, for a H$_2$O-rich atmosphere, the transit depth in the H$_2$O absorption bands increases linearly with temperature, since the atmospheric scale height, and hence the atmospheric thickness ($H_\lambda$), depend linearly on temperature, as discussed in Eq (\ref{eq:atmos1}) \& (\ref{eq:atmos3}) (also see Howe \& Burrows 2012). On the other hand, the transit depths in the opacity windows remain almost unaffected. Therefore, given a host star, a close-in super-Earth orbiting it offers better chances for atmospheric molecular detections compared to the same planet orbiting farther out. In this regard, even though cool super-Earths represent a natural progression towards finding habitable planets, they are less optimal for atmospheric characterization via transmission spectroscopy as exemplified by several recent studies of GJ~1214b (e.g. Bean et al. 2011; Berta et al. 2011; Desert et al. 2011). It is to be noted, however, as shown in the bottom panel of Fig.~\ref{fig:gj1214_2}, that the atmospheric thickness in scale heights ($N_{sc, \lambda}$) for a water-rich atmosphere is independent of the temperature, so Eq (\ref{eq:Nsc_lambda}) is still applicable. 

\subsubsection{Effect of Clouds}
\label{sec:clouds}

The presence of clouds in a super-Earth atmosphere can critically influence the interpretation of its transmission spectrum. Recent observations have suggested the possibility of clouds and/or hazes in a wide range of exoplanetary atmospheres (Pont et al. 2008; Madhusudhan et al. 2012; Marley et al. 2013; Morley et al. 2013). Clouds are particularly important in a super-Earth atmosphere because they can obscure the spectral features of the atmosphere leading to a nearly featureless `flat' transmission spectrum in the infrared. A similar spectrum with subsided spectral features can also be caused due to a high mean-molecular mass, e.g. rich in H$_2$O, CO$_2$, etc., as shown in Fig.~\ref{fig:gj1214_1}. Due to this degenerate set of solutions, observations of featureless spectra can be challenging to interpret, requiring very high precision observations to break the degeneracy. The super-Earth GJ~1214b is a classic example in this regard (discussed in detail in section~\ref{sec:gj1214b}). %

High-temperature super-Earths are better candidates for atmospheric characterization due to the lower probability of clouds in their atmospheres. The presence of clouds, and their chemical composition, is a strong function of the temperature. Several recent studies have investigated the compositions of clouds in exoplanetary atmospheres (e.g. Kempton et al. 2012; Howe \& Burrows 2012; Marley et al. 2013; Morley et al. 2013). Figure~\ref{fig:snrplots} shows condensation temperatures for several compounds expected in planetary atmospheres. At very low temperatures (T $\lesssim$ 300 K), H$_2$O itself condenses out of the upper atmosphere, making spectroscopic observations of H$_2$O in super-Earths extremely challenging, similar to the challenges in measuring H$_2$O abundances in giant planets in the solar system (see e.g. Atreya 2010). On the other hand, even for warmer atmospheres (T $\sim$ 300 - 1000 K), several other volatile species, such as NaCl, KCl, Na$_2$S, etc., can condense out leading to cloudy super-Earth atmospheres in this temperature range, as is likely the case with the super-Earth GJ~1214b (Bean et al. 2011; Morley et al. 2013). Furthermore, even very high temperature atmospheres (T $\sim$ 1000 - 2000 K) can host clouds made of refractory species (e.g. MgSiO$_3$, Fe, etc.). 

Refractory condensates which form in high-T atmospheres tend to be heavier than low-T volatile condensates, and, hence, would likely lead to less extended cloud altitudes due to efficient gravitational settling of the condensates (see e.g. Spiegel et al. 2009). Consequently, higher-T super-Earth atmospheres could be expected to have lesser cloud covers, leading to strong spectral signatures. Furthermore, extremely irradiated super-Earths with T $\gtrsim$ 2000 K, such as 55~Cancri~e, form a potentially ideal sample with `cloud free' atmospheres which present the best chances for detecting H$_2$O-features in super-Earth spectra. 

\subsection{Thermal Emission Spectra}
\label{sec:emission}

Contrary to transmission spectra, thermal emission spectra of a transiting planet observed at secondary eclipse probe its dayside atmosphere. Such spectra allow constraints on both the chemical composition as well as temperature profile of the planet's dayside atmosphere. Observations of thermal emission have been reported for several dozens of giant exoplanets, but are only beginning for transiting super-Earths (e.g. Demory et al. 2012; Gillon et al. 2013). The primary challenge is that the eclipse depth, which is a measure of the planet-star flux ratio, depends on both the radius and temperature of the planet and star, and can be significantly smaller than the transit depth. The planet-star flux ratio is given by: 

\begin{equation}
\frac{f_{p\lambda}}{f_{s\lambda}} = \frac{B_\lambda(T_p) R_p^2}{B_\lambda(T_s)R_s^2}  
\label{eq:emission}
\end{equation}
where $f_{p\lambda}$ and $f_{s\lambda}$ are the fluxes from the planet and star, respectively, $B_\lambda(T)$ is the Planck function, and $T_p$ ($T_s$) and $R_p$ ($R_s$) are the brightness temperature and radius of the planet (star), respectively. As evident from Eq~(\ref{eq:emission}), the eclipse depth is lower than the transit depth, Eq~(\ref{eq:atmos2}), by a factor of $B_\lambda(T_p)/B_\lambda(T_s)$, and is therefore harder to detect, especially for cooler planets. It is clear from Eq~(\ref{eq:emission}) that for a given star, bigger and hotter planets lead to higher planet-star flux contrasts. Consequently, the only robust detection of thermal emission from a super-Earth has been reported for 55 Cancri e with $T_{\rm eq} \sim 2000-2400$ K (Demory et al. 2012). Despite the apparent challenge in observing thermal emission from super-Earths, emission spectra provide unique constraints on the vertical temperature profile and chemical composition on the dayside atmosphere which is inaccessible from transmission spectra (see section~\ref{sec:case_studies}). 

\subsubsection{`Surface Temperature' and Atmospheric Constraints}
\label{sec:surface_temp}

Observations of thermal emission from super-Earths in opacity windows allow determination of their `surface' temperatures ($T_{sf}$). As discussed in section~\ref{sec:surface_radius}, opacity windows are wavelengths where the opacity in the planetary atmosphere is minimal. Radiation emitted from the planetary surface in these spectral bandpasses traverse unimpeded through the planetary atmosphere before reaching the observer. Consequently, brightness temperatures measured in such bandpasses constrain the `surface temperatures' of the super-Earths. For planets with gaseous atmospheres, such observations can constrain the temperature in the lower atmosphere (Madhusudhan 2012), irrespective of the atmospheric composition. We discuss these aspects in section~\ref{sec:55Cnc-emission} and Fig.~\ref{fig:55Cnc_emission}. For example, Fig.~\ref{fig:55Cnc_emission} shows model emission spectra for super-Earth 55 Cancri e with different atmospheric compositions (solar composition and H$_2$O-rich), and a blackbody spectrum corresponding to the temperature of the lower atmosphere, i.e. the surface temperature. As shown in the figure, the different spectra converge to the blackbody spectrum in the opacity windows in the $Y$, $J$, $H$, and, to some extent, the $K$ bands. 

On the other hand, for wavelengths corresponding to molecular absorption bands, the $\tau \sim 1$ surface lies higher up in the atmosphere. Therefore, brightness temperatures measured in the molecular bands allow joint constraints on the temperature profiles and molecular composition of the dayside atmosphere, as has been extensively demonstrated for hot Jupiter atmospheres (Madhusudhan et al. 2011). We discuss model atmospheres and spectral features in thermal emission for specific super-Earths in section~\ref{sec:case_studies}. 

\begin{figure}
\includegraphics[width = 0.5 \textwidth]{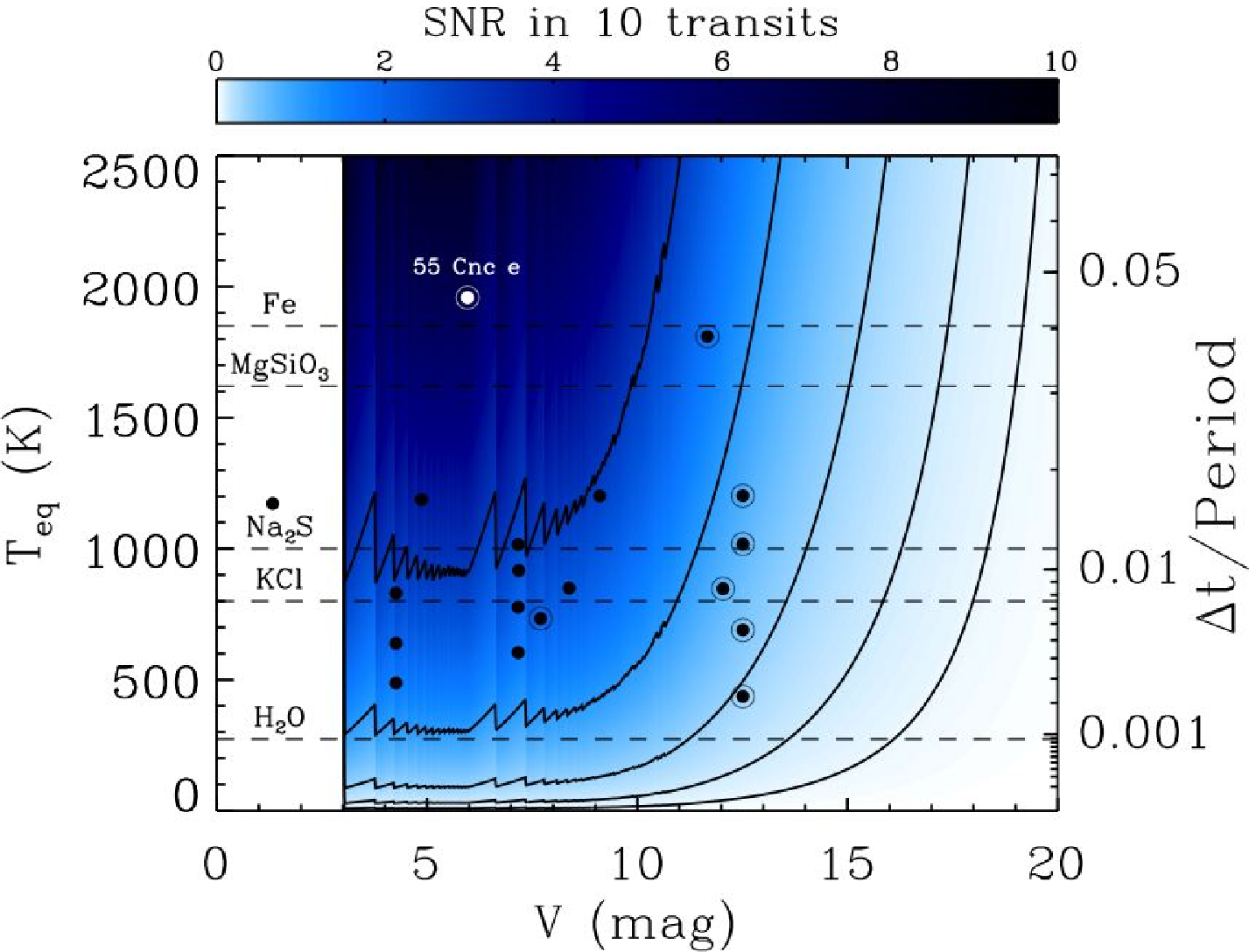}
\includegraphics[width = 0.5 \textwidth]{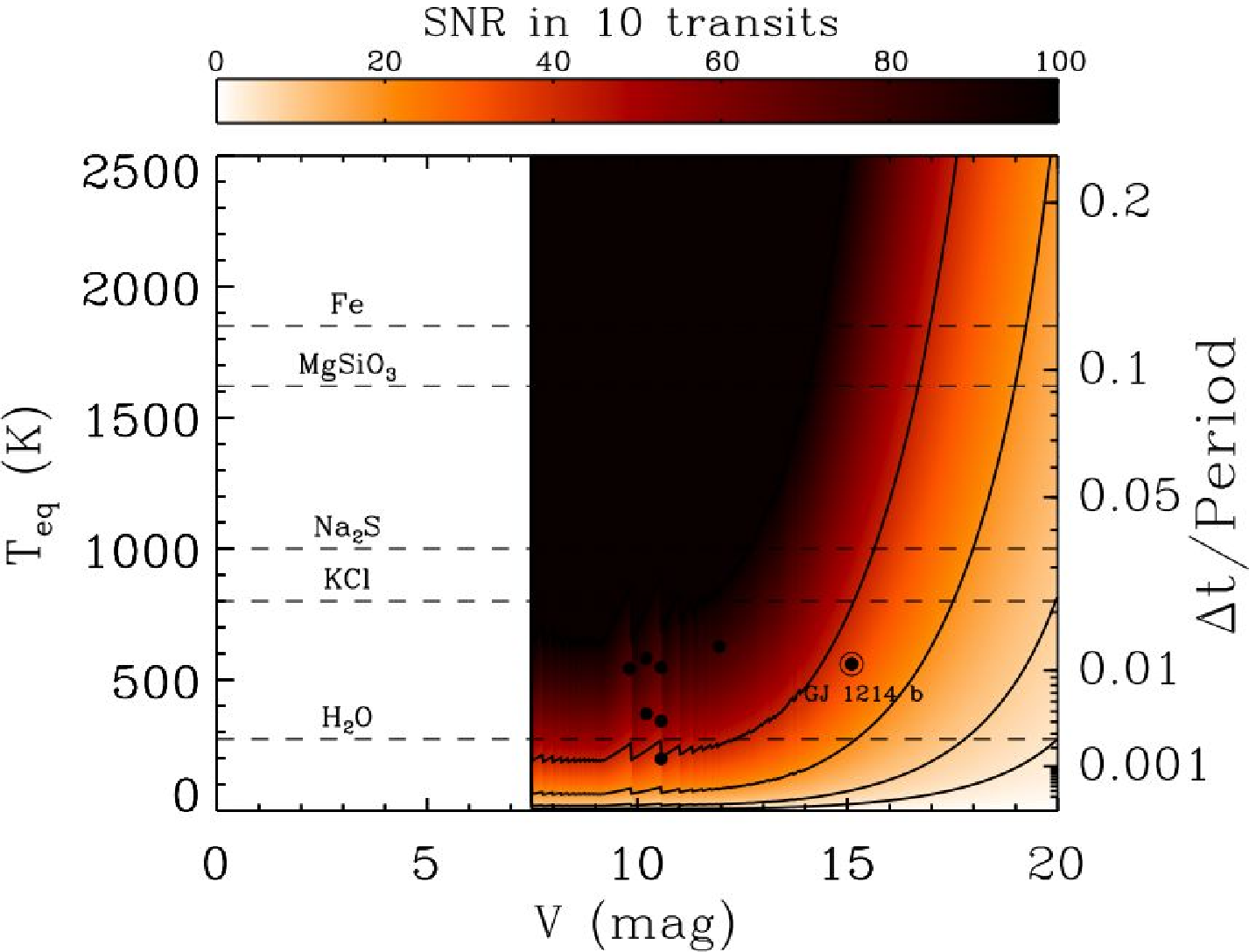}
\caption{Sensitivity simulations of the detection of the atmospheric H$_2$O feature at 1.4 $\mu$m in H$_2$O-rich super-Earths using the {\it HST}  WFC3 spectrograph in spatial scanning mode. Calculations were made for transits of a G8V star (e.g., 55 Cnc) in the top panel (blue), and transits of an M4V star (e.g., GJ 1214) in the bottom panel (red).  S/N contours of 0.03, 0.1, 0.3, 1, and 3 are shown in top panel, and contours of 1, 3, 10, and 30 are shown in the bottom panel.  In 10 transits, H$_2$O would be detected in 55 Cnc e at a S/N $\sim$ 6.4 and GJ 1214b a S/N $\sim$ 21.5.  Parameter space in the GJ 1214 panel is limited to stars with $V \geq 7.5$, set by the brightest known M star (GJ 411).  Other known exoplanetary systems with similar host star spectral type are shown with filled circles, and transiting exoplanets indicated with large open circles. Condensation curves of various compounds at 1 bar are shown.  Sensitivities are calculated assuming no obscuring clouds.  In scenarios where clouds are present, a definitive detection of the atmosphere may not be possible, even at high S/N, e.g., in the case of GJ~1214b.}
\label{fig:snrplots}
\end{figure}

\section{Prospects with HST and Optimal Discovery Space}
\label{sec:HST+discovery}

In this section, we investigate the following fundamental question: What properties of 
super-Earths and their host stars allow the best chances to detect H$_2$O in their 
atmospheres with existing facilities? To this end, we consider {\it HST} Wide Field Camera 3 (WFC3) spectrograph 
as our instrument of choice, since this is the only space-based instrument which currently has the 
spectroscopic capability to detect H$_2$O in exoplanetary atmospheres (e.g. Deming et al. 2013). 
In our study, we consider two archetypes for stellar hosts and planetary sizes, GJ~1214b and 55 Cancri e, 
and estimate the detectability of H$_2$O features in each case but over a wide range in planetary 
temperatures and stellar brightnesses. 

Our sensitivity estimates, shown in Fig.~\ref{fig:snrplots} and described below, were
derived in the context of a detection of the 1.4~$\mu$m H$_2$O
feature.  However, the results are applicable to any feature in the
infrared passband from 0.9--1.65~$\mu$m.  The recently implemented
spatial scanning capability of {\it HST} WFC3 makes it
possible to observe bright targets with a dense temporal sampling.
This technique has proved very successful for exoplanet transmission
spectroscopy (e.g., McCullough \& MacKenty 2012, Deming
et al. 2013).  The efficiency and precision of the spectrophotometry
are greatly improved by scanning the point source over the detector
such that the counts are distributed over a wide range of pixels.

While the current Exposure Time Calculator (ETC) for WFC3 does not
calculate the implementation of this new scanning mode, it is
straightforward to estimate the exposure time and expected S/N, as
detailed in McCullough \& MacKenty (2012). We assume a 256 pixel
subarray to ensure parallel buffer dumps so that there
is no interruption in the data acquisition.  We use model spectra of a
G8V and an M4V spectral class (Pickles 1998), scaled to the
appropriate brightness, to estimate the count rate in the passband of
interest as a function of $H$ magnitude, which is the band close to the 
1.4~$\mu$m H$_2$O line of interest. The count contributions of the sky, 
dark current, thermal background, and readout noise derived using the ETC 
are minimal given the brightnesses of targets in question.  

Instead of the light being concentrated as a point source, it is
distributed over scores of pixels.  However, the scan rate ($R$ in
arcsec s$^{-1}$) is limited to 5 arcsec s$^{-1}$. The maximum number of
scanned pixels is 200, although less are used if the count rate
permits and a lower scan rate is needed. In order to avoid approaching the 
nonlinearity regime of the detector, we limit the exposure to $<$60\% of the 
maximum well depth. A nominal {\it HST} orbit length of 2700 seconds
is assumed.  Together with the optimal exposure time, we include the
nominal overhead time (60 seconds), scan return time (5 seconds), and
the readout time (0.3 seconds). 

The S/N estimates are based on the expected atmospheric signal as described in Eq (\ref{eq:atmos3}). The noise estimates are derived by binning over the entire 1.4~$\mu$m H$_2$O feature using a passband of 0.2~$\mu$m. Sensitivity estimates are made for two scenarios, 
one corresponding to circumstances similar to 55 Cnc e, a super-Earth in 
orbit around a G8V star (top panel in Fig.~\ref{fig:snrplots}), and one similar to GJ
1214b, a super-Earth in orbit around an M star (bottom panel in
Fig.~\ref{fig:snrplots}).  Our sensitivity simulations match the actual noise levels of the first observations of transiting exoplanets using the spatial scanning mode (Deming et al. 2013).  

In calculating the signal for each scenario, we adopt the planetary radius and gravity and the stellar 
spectrum of the prototype, and we vary the equilibrium temperature (i.e., semimajor axis)
of the planet and the stellar brightness. Both planets, GJ 1214b and 55 Cnc e, are identified, and shown with the sample of known super-Earth (i.e., $M\sin i < 10$) exoplanets with host stars
that have the same approximate spectral type (4900 K $< T_{\rm eff} <$
5600 K, roughly G5 to K2 for the 55 Cnc example, and $T_{\rm eff} <
3800$ K, roughly any M star for the GJ 1214 example).  Those
exoplanets that are known to transit are shown with a large open circle
surrounding the filled symbol.  In the case of GJ 1214, it is the only
known transiting exoplanet that fits these criteria.  

Ostensibly, our estimates agree with the common notion that brighter and/or smaller 
host stars greatly enhance the detectability of super-Earth atmospheres. While a higher 
planetary temperature around a larger star, as in the case of 55 Cnc e, can also 
enhance the transit signal, the dependence of the signal on the stellar 
radius is stronger, as shown in Eq (\ref{eq:atmos3}). For example, in these calculations, 
we estimate a S/N $\sim 21.5$ in the 1.4 $\mu$m H$_2$O feature to be obtained in observations 
of ten transits of GJ 1214b (S/N $=$ 1, 3, 10, and 30 contours are shown), whereas for 55 Cnc e, 
ten transits results in an estimated S/N of $\sim$ 6.4 (S/N $=$ 0.03, 0.1, 0.3, 1, and 3 contours are shown). 

However, there are two additional important factors, beyond the nominal sensitivity 
estimates discussed above, that need to be considered while planning atmospheric 
observations of super-Earths. Firstly, as discussed in section~\ref{sec:clouds}, clouds 
can play a critical role in low temperature atmospheres of super-Earths. In our sensitivity 
estimates discuss above, we have assumed spectral features of H$_2$O as observed in a cloud-free 
atmosphere. However, the strength of the actual atmospheric signal is highly dependent 
on the presence or absence of clouds. Fig.~\ref{fig:snrplots} shows condensation temperatures 
of some common species at a nominal 1-bar pressure (Lodders 2002; Sudarsky et al. 2003; 
Morley et al. 2013). As shown in the figure, a wide range of condensates are possible in super-Earth 
atmospheres, volatile species for temperatures below $\sim$1000 K and refractory species for 
temperatures as high as $\sim$2000 K. The presence of resulting clouds, if present at high enough 
altitudes, can even completely mask the spectral features in a transmission spectrum leading to a 
featureless spectrum. 

Consequently, atmospheres of low-temperature super-Earths (T$_{\rm eq} \lesssim 1000$) can be challenging 
to observe and interpret even if the planet-star radius ratios and sensitivity estimates are highly favorable, 
as is the case for GJ~1214b, as discussed in section~\ref{sec:gj1214b}. While super-Earths with T$_{\rm eq} \sim 1000 - 2000$ can still host clouds made of refractory condensates (silicates, Fe, etc.) the cloud-altitude would likely be lower compared to volatile condensates (e.g. NaCl, H$_2$O, etc.) owing to more efficient gravitational settling due to heavier molecules. On the other hand, very high-$T$ super-Earths with T$_{\rm eq} \gtrsim 2000$, such as 55 Cancri e, likely present the clearest atmospheres with the best potential for observations of their spectra. The super-Earth 55 Cancri e is particularly favorable given its extremely bright ($V = 5.95$; $H = 4.27$) host star. 

The second factor is that the short periods of high-T$_{\rm eq}$ super-Earths provide a much higher number of transit opportunities. Atmospheric characterization of any super-Earth with current instruments, such as the {\it HST} WFC3, require co-adding multiple transits to be able to make definitive detections of molecular features. Consequently, it is important to be able to schedule multiple transits for a dedicated program to characterize super-Earth atmospheres. Short period planets are advantageous in this regard by offering a higher frequency of transit opportunities. For example, for 55 Cnc e, $\sim$495 transits occur in a year, compared to 230 transits of GJ 1214b in a year.  

\section{Case Studies}
\label{sec:case_studies}

The small sizes of super-Earths means that observing their atmospheres require particularly favorable conditions that either enhance the signal or improve the precision. For a given planet size, the transit and eclipse depths are larger for smaller host stars and hotter planets. On the other hand, a brighter host star yields better precision in the observations. While GJ~1214b orbits a small star (an M dwarf), 55 Cancri e orbits an extremely bright (V=6) sun-like star in a very close (18-hour) orbit, because of which both these super-Earths are great candidates for atmospheric studies. In what follows, we discuss the atmospheric constraints for these two planets possible with observations using existing facilities. 

\subsection{The Atmosphere of super-Earth GJ 1214b}
\label{sec:gj1214b}

The super-Earth GJ~1214b (Charbonneau et al. 2009) is one of the most observed exoplanets to date. The planet, with a mass (M$_p$) of 6.47 M$_\oplus$ and a radius (R$_p$) of 2.68 R$_\oplus$ orbits a late M dwarf (R$_s$ = 0.21 R$\odot$, T$_{\rm eff}$ = 3030 K), resulting in a large transit depth. Consequently, despite the relatively faint host star (V = 15.1) and a relatively low equilibrium temperature (T$_{\rm eq} \sim$ 550 K), the planet is particularly suitable for transit observations and atmospheric characterization. 

\begin{figure}[t]
\centering
\includegraphics[width = 0.5\textwidth]{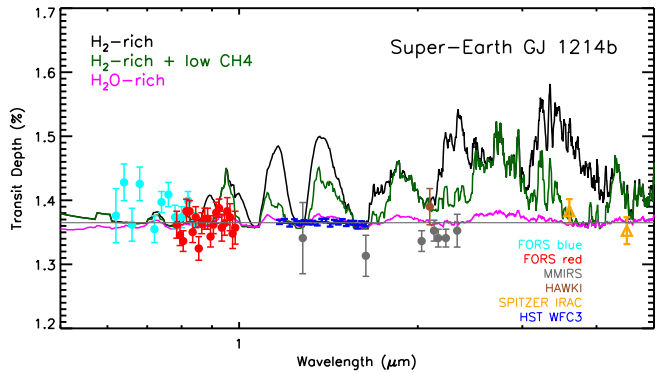}
\caption{Observations and model transmission spectra of GJ1214b. The solid curves show model spectra of GJ~1214b with the different chemical compositions  described in the legend. The black (green) curve correspond to a H$_2$-rich atmosphere model assuming solar abundances and chemical equilibrium (non-equilibrium). In the green model CH$_4$ is depleted by a factor of 100 relative to chemical equilibrium, motivated by similar requirements for the hot Neptune GJ~436b (Madhusudhan \& Seager 2011). The magenta model corresponds to a H$_2$O-rich atmosphere of GJ~1214b. The various symbols with error bars show the observations using different instruments from space and ground, as described in the legend and reported by Bean et al. (2011), Desert et al. (2011) and Kreidberg et al. (2014). Only a subset of all available observations in the literature are shown here for clarity. All the observations in the literature to date are consistent with a flat spectrum, shown as a gray horizontal line, but are also consistent with a cloudy atmosphere of unknown composition. The observations, however, rule out a cloud-free H$_2$-rich or H$_2$O-rich atmosphere (also see Kriedberg et al. 2014). In the present work, we also suggest that a H$_2$O-rich atmosphere is unlikely based on internal structure models (see section~\ref{sec:gj1214_interior}.)} 
\label{fig:gj1214_trans}
\end{figure}

\subsubsection{Constraints from Transmission Spectroscopy}
\label{sec:gj1214_trans}

Transmission spectroscopy and photometry of the planet's atmosphere at the day-night terminator have been reported over a wide spectral baseline ranging from the visible to mid-infrared ($\sim 0.6 - 5~\mu$m) (e.g. Bean et al. 2010,2011; Berta et al. 2011; Croll et al. 2011; Desert et al. 2011; de Mooij et al. 2011; Kriedberg et al. 2014). The observed bandpasses, from a wide array of facilities from ground and space, encompass spectral features of prominent molecules such as H$_2$O and CH$_2$, as well as continuum bandpasses or opacity windows (see section~\ref{sec:surface_radius}). Several modeling efforts in the recent past have aided in the interpretation of the observed spectra (see e.g. Miller-Ricci et al. 2011; Bean et al. 2011; Miller-Ricci \& Fortney 2010; Benneke \& Seager 2012; Kepmton et al. 2012; Howe \& Burrows 2012; Morley et al. 2013). 

The sum total of data show that the transmission spectrum of the planet is consistent with a flat horizontal line over the entire wavelength range observed to date, as shown in Fig.~\ref{fig:gj1214_trans}. The flat transmission spectrum of GJ~1214b rules out a cloud-free atmosphere for several plausible compositions, e.g. dominated by H$_2$, H$_2$O, CO$_2$, or CH$_4$ (Kreidberg et al. 2014). However, a featureless transmission spectrum is also consistent with a cloudy atmosphere, of unconstrained composition including a solar abundance H$_2$-rich composition. In this scenario, the presence of  clouds aloft in the atmosphere could be blocking the starlight thereby masking out any molecular absorption features due to the planetary atmosphere. Currently available data are, therefore, inconclusive about the true atmospheric composition of GJ~1214b. 

\subsubsection{Constraints from Internal Structure Models}
\label{sec:gj1214_interior}

The observed mass and radius of GJ~1214b, together with internal structure models, suggest that a cloud-free H$_2$O-rich atmospheric composition is unlikely for this planet. As discussed in section~\ref{sec:interior}, GJ~1214b can be classified as an SE3 type super-Earth with its radius, $R_p$, being larger the radius ($R_{\rm H_2O}$) of a pure-H$_2$O planet with the same mass. As such, the radius differential  between $R_p$ and $R_{\rm H_2O}$ can be used to place an upper limit on the mean molecular mass of the planetary atmosphere, as given by Eq (\ref{eq:rp2}). Considering the parameters of GJ~1214b, we find a $\mu_{max}$ of 2.0 $\pm$ 0.7 amu, consistent with a H$_2$-rich atmosphere.  Consequently, it is less likely that the observed radius of GJ~1214b can be explained solely by a cloud-free atmosphere with a high $\mu$, e.g. H$_2$O-rich, in agreement with what is already known from the transmission spectrum as discussed above. 

A low atmospheric $\mu$ for GJ~1214b would also be consistent with internal structure models which suggest a light-element (H/He) composition for the planetary envelope (Rogers \& Seager 2010; Valencia et al. 2013). As shown in Fig.~\ref{fig:m-r}, the radius of the planet is higher than the $R_{\rm p,max}$ for a pure-H$_2$O planet with a H$_2$O-rich atmosphere for the mass of GJ~1214b and its $T_{\rm eq}$ of $\sim550$K. We find that a purely H$_2$O-ice interior of GJ~1214b would require $N_{sc} \gtrsim 75$ scale heights of a H$_2$O-rich atmosphere to explain the radius, which is physically implausible, and an even higher $N_{sc}$ for other gases such as CO$_2$ or N$_2$. Consequently, as discussed above, a significantly lighter element than H$_2$O, such as a H-rich atmosphere would be required to explain the radius. Our interpretation is consistent with the results of Rogers \& Seager (2010) who also suggested the requirement of a H/He envelope in the planet to explain its mass and radius. While our results rule out a cloud-free H$_2$O-rich atmosphere in GJ~1214b, a H$_2$O-rich lower atmosphere with a lighter species in the upper atmosphere together with a very high-altitude cloud/haze cover, cannot be conclusively ruled out in the present work. 

\subsubsection{Constraints from Thermal Emission Spectra} 
\label{fig:gj1214_thermal}

Observations of thermal emission from GJ~1214b have led to only nominal constraints on its atmospheric composition. Given the low temperature of the planet, thermal emission from GJ~1214b is challenging to observe with existing instruments. Recently, Gillon et al. (2013b) reported upper-limits on thermal emission in the {\it Spitzer} photometric bands at 3.6 $\mu$m and 4.5 $\mu$m. The data are consistent with conclusions derived from transmission spectroscopy of GJ~1214b, as discussed in section~\ref{sec:gj1214_trans}. The data rule out a cloud-free H$_2$-rich composition in the dayside atmosphere of the planet. However, the data are consistent with a H$_2$O-rich atmosphere as well as a cloudy H$_2$-rich atmosphere. Additionally, the data are also consistent with a blackbody spectrum with a temperature of 500-600 K, indicating the possibility of an isothermal temperature structure with unconstrained chemical composition. Consequently, current observations of thermal emission from GJ~1214b do not provide any significant constraints beyond what is already known from extensive observations of transmission spectra of the planet. 

New observations of thermal emission from GJ~1214b with existing facilities will be challenging, if not impossible. In the wavelength range ($\sim$ 1 -- 2.3 $\mu$m) of current instruments in the near-infrared, e.g. the {\it HST} WFC3 spectrograph and ground-based instruments, the predicted planet-star flux contrast in dayside thermal emission is below 40 ppm. Detecting such a signal, would require precisions better than 10 ppm which current instruments are not likely to achieve, particularly given the faint host star (V = 15.1). In the future, however, the James Webb Space Telescope ({\it JWST}) will be able to detect such a weak signal. 

\subsection{The Atmosphere of super-Earth 55 Cancri e}
\label{sec:55Cnc_hst}

The super-Earth 55 Cancri e presents arguably the best chance for comprehensively characterizing a super-Earth atmosphere using both transmission as well as thermal emission spectroscopy. The planet has a mass of 8.4 $M_\oplus$ and a visible radius of 2.0 $R_\oplus$, and orbits a nearby G dwarf at a period of 18 hours (Demory et al. 2011; Winn et al. 2011; Endl et al. 2012). The parent star is the brightest star (V = 6) known to host a transiting exoplanet, and has led to measurements of the planet's radius at exquisite precision in the visible as well as in the Spitzer 4.5 $\mu$m IRAC band (Winn et al. 2011; Demory et al. 2012; Gillon et al. 2012). Furthermore, due to its very  short orbit, the planet has an equilibrium temperature of  $\sim$2000-2400 K, which leads to significant thermal emission, as has been observed in the Spitzer 4.5 $\mu$m IRAC photometric band, the first for any super-Earth (Demory et al. 2012). 

\subsubsection{Constraints from Internal Structure Models}
\label{sec:55cnc_interior}
55 Cancri e is an SE2-type super-Earth, as described in section~\ref{sec:interior}, with its $M_p$ and $R_p$ lying between the mass-radius relations of a pure-silicate and a pure-H$_2$O planet, as shown in Fig.~\ref{fig:m-r}. As such, the existence of a potential atmosphere cannot be conclusively constrained without multi-color atmospheric observations (discussed in section \ref{sec:55Cnc-trans}). However, the $M_p$ and $R_p$ of 55 Cancri e have led to two contrasting hypotheses for its interior composition, with different implications for its atmospheric composition. Considering a terrestrial-like oxygen-rich mineralogy, consisting of Fe, silicates, and H$_2$O, would require that the planet host a massive ($\gtrsim$ 10 \%) envelope of supercritical H$_2$O (Valencia et al. 2010) in order to explain the $R_p$, as suggested in several recent works (Winn et al. 2011; Demory e et al. 2011,2012; Gillon et al. 2012). However, it is yet to be  conclusively demonstrated if a massive H$_2$O envelope of the planet would be stable against atmospheric escape (Valencia et al. 2010) and instability from night-side condensation (Castan \& Menou 2011; Heng \& Kopparla 2012) given the long age (10.2 $\pm$ 2.5 Gyr; von Braun et al. 2012) and extreme irradiation of the system. An alternate hypothesis suggests that the planet is carbon-rich, composed of Fe, C (as graphite+diamond), SiC, and silicates, without the requirement of any volatile envelope (Madhusudhan et al. 2012). 

The contrasting interpretations for the interior composition of 55 Cancri e suggest three possible compositions for its atmosphere. Firstly, an oxygen-rich composition in the planet requires that the planet host a massive H$_2$O envelope causing a H$_2$O-rich atmosphere. Ehrenreich et al. (2012) place an upper-limit on the escape rate of hydrogen resulting from photodissociation of a water-rich atmosphere, and find it consistent with a stable H$_2$O-rich atmosphere. Secondly, a carbon-rich composition would be unlikely to host a H$_2$O-rich atmosphere, and may even host no atmosphere at all. Finally, the data can also be explained with a H$_2$-rich atmosphere overlying an interior of any composition, oxygen-rich or carbon-rich. These three scenarios for the atmospheric composition of 55 Cancri e can be constrained using spectroscopic observations as discussed below. 

\subsubsection{Constraints from Transmission Spectra} 
\label{sec:55Cnc-trans}

\begin{figure}
\includegraphics[width = 0.5\textwidth]{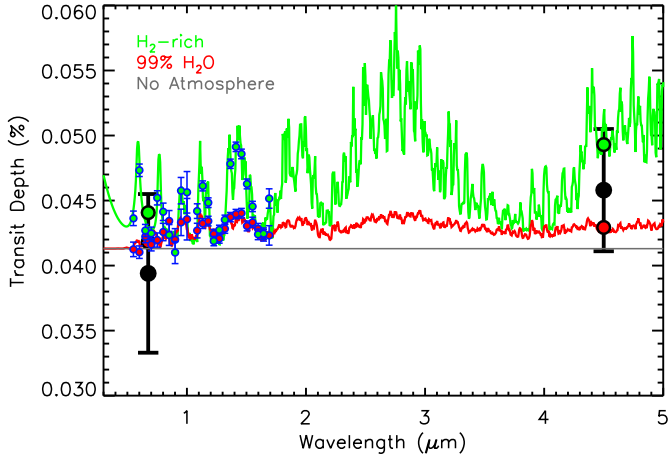}
\caption{Model transmission spectra of super-Earth 55 Cancri e. The green and red solid curves show two model spectra of the planet with H$_2$-rich and H$_2$O-rich compositions, respectively. The gray horizontal curve shows  a model spectrum with no atmosphere. The large black circles with uncertainties show photometric measurements of the transit depth in the visible, using the {\it MOST} telescope (Winn et al. 2011), and in the infrared at 4.5 $\mu$m using the {\it Spitzer} space telescope (Gillon et al. 2012). The small blue circles with uncertainties show our simulated observations with WFC3 in spatial scanning mode for both the model atmospheres. {\it HST} WFC3 observations would be able to provide an unambiguous constraint on the chemical composition of the atmosphere, particularly on the possibility of a H$_2$O-rich atmosphere. The high temperatures on the planet ($T \sim 2000 - 2400$ K) also imply that the atmosphere is very likely cloud-free, thereby removing ambiguities in interpretation of the spectra.}
\label{fig:55Cnc_trans}
\end{figure}

Transmission spectroscopy of 55 Cancri e using {\it HST} WFC3 can constrain the atmospheric composition at the day-night terminator of the planet. Fig.~\ref{fig:55Cnc_trans} shows our model transmission spectra of 55 Cancri e in the three possible atmospheric scenarios (H$_2$-rich, H$_2$O-rich, and no atmosphere) and simulated observations assuming 10 transits observed with the {\it HST} WFC3 as derived in section~\ref{sec:HST+discovery}. As shown in Fig.~\ref{fig:55Cnc_trans}, the radii of 55 Cancri e previously measured in photometric bandpasses in the visible and Spitzer IRAC 4.5 $\mu$m band are consistent with any of the three possible compositions. On the other hand, WFC3 observations will be able to conclusively constrain all the three scenarios. As discussed in section~\ref{sec:gj1214b}, previous studies have attempted to make a similar determination using {\it HST} WFC3 for the super-Earth GJ~1214b. However, due to the much lower temperature of GJ~1214b ($\sim$500 K) clouds of various compositions are possible (e.g. Morley et al. 2013), making a conclusive determination of a H$_2$O-rich atmosphere difficult for that planet. On the other hand, at the 2000-2400 K temperature of 55 Cancri e, clouds of any known composition are highly unlikely to exist, thereby making the interpretation of its transmission spectra substantially easier than that of GJ~1214b.

{\it HST} observations will be able to detect the presence of a H$_2$O-rich atmosphere better than 6-$\sigma$ in the water bands in the WFC3 bandpass, and the presence of a H$_2$-rich atmosphere at better than 10-$\sigma$ in the same bands. It is important to note that a non-detection in the WFC3 bandpass, though highly unlikely, would be an extremely important result. Such a result would strongly imply the lack of an atmosphere in 55 Cancri e, for which the only known explanation to date is one of a carbon-rich interior, as discussed in section~\ref{sec:55cnc_interior}. Consequently, a flat spectrum across multi-wavelength observations, including WFC3, of 55 Cancri e can provide conclusive evidence for the lack of a H$_2$-rich or H$_2$O-rich atmosphere in the planet, without the ambiguities that plague similar efforts for GJ~1214b. 

\begin{figure}
\includegraphics[width = 0.45\textwidth]{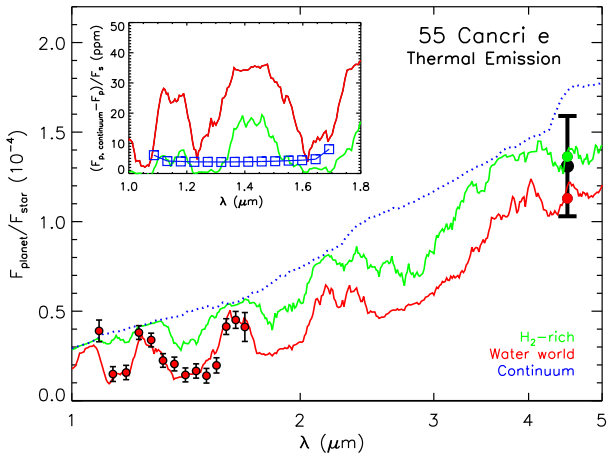}
\caption{Thermal emission spectra for 55 Cancri e predicted for different atmospheric compositions.  The red curve indicates a H$_2$O-rich atmosphere, while the green curve is representative of an H$_2$-rich atmosphere.  The blue curve shows a blackbody curve corresponding to the bottom optically thick `surface' of the atmosphere, if present. Red data points indicate our simulated WFC3 secondary eclipse observations taken using spatial scanning (see section~\ref{sec:55Cnc-emission}). The scatter and error bars are shown assuming a water-rich atmosphere. The large black circle with uncertainties shows the {\it Spitzer} 4.5 $\mu$m detection (Demory et al. 2012).  The inset shows the a close-up of the WFC3 IR red spectral region. The curves show the H$_2$O-rich and H$_2$-rich models relative to the blackbody continuum, and the blue squares show the predicted sensitivities. {\it HST} WFC3 observations will be capable of detecting thermal emission from the planet in all the modeled scenarios and will be able to discriminate between the different models and detect a H$_2$O-rich atmosphere if present, as discussed in section~\ref{sec:55Cnc-emission}.}
\label{fig:55Cnc_emission}
\end{figure}

\subsubsection{Constraints from Thermal Emission Spectra}
\label{sec:55Cnc-emission}

55 Cancri e is the only super-Earth for which thermal emission from the planet can be detected using existing instruments, given the extremely high dayside temperature (2400 K) of the planet. Figure~\ref{fig:55Cnc_emission} shows model thermal emission spectra of 55 Cancri e for the two possible atmospheric compositions, i.e. H$_2$O-rich versus H$_2$-rich based on solar abundances. We find that a precision of $\sim 5$ ppm, which is attainable with ten eclipses observed with {\it HST} WFC3, would be able to detect thermal emission in the opacity windows, i.e. from the planetary `surface' (see section \ref{sec:surface_temp}), at $\gtrsim$ 5-$\sigma$. The H$_2$O feature in either scenario, which is given by the flux differential between the blackbody continuum, in the opacity windows, and the emission within the water band, can be detected at $\gtrsim$ 4-$\sigma$ for both scenarios. A non-detection of an H$_2$O feature, i.e. the observation of a blackbody spectrum,  would imply either the lack of an atmosphere, or an isothermal temperature profile. The degeneracy between the two solutions can be lifted by complementary observations of transmission spectra as discussed in section~\ref{sec:55Cnc-trans}. 

A thermal emission spectrum of 55 Cancri e obtained with {\it HST} WFC3 can allow three specific constraints that are unprecedented for a super-Earth. Firstly, one will be able to determine the thermal profile of the dayside atmosphere of the planet. A thermal profile decreasing with altitude will give rise to molecular absorption features, as seen in the deep water features in Fig.~\ref{fig:55Cnc_emission}. Secondly, the difference between the continuum regions and strong water absorption or emission features in the WFC3 bandpass can be used to place joint constraints on the temperature gradient as well as the H$_2$O-abundance in the atmosphere using detailed retrieval algorithms (e.g. Madhusudhan \& Seager 2009; Madhusudhan et al. 2011; Lee et al. 2012; Line et al. 2012). Thirdly, the spectral energy distribution of thermal emission can be used to constrain the day-night energy redistribution in the planetary atmosphere (Madhusudhan \& Seager 2009).

\section{Discussion and Summary} 
\label{sec:discussion}

 In the present work, we find that detection and characterization of hot super-Earths orbiting nearby stars provide a greater strategic advantage in the coming decade than detecting cool super-Earths close to the habitable zones of their host stars. A habitable planet, with temperatures in the vicinity of $\sim$300 K to sustain liquid water on the surface, is not necessarily inhabited. At a minimum, such a determination would require characterizing the planet's atmospheric composition. However, the atmospheres of such cool planets will be challenging to detect even using the best facilities expected in the coming decade, including the James Webb Space Telescope (\textit{JWST}). Even when detected, the likely presence of clouds will confound the interpretation of spectra, as is already the case for the super-Earths GJ~1214b (with $T \sim 500$ K) even with the most extensive observations with current facilities, including {\it HST}, {\it Spitzer}, and major ground-based facilities. In the future, however, {\it JWST} is expected to be able to characterize super-Earths like GJ~1214b to greater precision, and more so for hotter planets which will have stronger atmospheric signatures, as discussed in the present work. 
 
  At the present time, a more important and tractable question than characterizing habitable exoplanets, is whether H$_2$O is indeed a dominant constituent of super-Earth atmospheres. Addressing this question is independent of whether the planet is habitable or not. In fact, the best super-Earths that would be able to help answer this question would be transiting super-Earths with very high temperatures and orbit host stars that are either very bright, to allow better spectroscopic precision, or are small in size, to allow larger planet-star radius ratio and flux ratio. Currently, three super-Earths are known to transit such favorable host stars, albeit with a wide range of temperatures: GJ~1214b (T$_{\rm eq}$ $\sim$ 550 K; Charbonneau et al. 2009), HD~976548 (T$_{eq}$ $\sim$ 900 K; Dragomir et al. 2013), and 55 Cancri e (T$_{eq}$ $\sim$ 2000-2400 K; Winn et al. 2011). Of these, 55 Cancri e offers the best chances for detecting H$_2$O in both transmission and thermal emission as discussed in the present work. Such hot super-Earths form an ideal sample in which to investigate the possibility and chemical abundances of H$_2$O-rich atmospheres in super-Earths. 
  
 Current and upcoming observational facilities are well suited to characterize H$_2$O-rich atmospheres in hot transiting super-Earths. We demonstrate in the present work how multi-color observations of H$_2$O-rich super-Earths can constrain the surface radii and temperatures of super-Earths, as well as constrain their H$_2$O abundances. Observations in the opacity windows, which probe the lower region (`surface') of a super-Earth atmosphere, are possible with a wide range of ground-based facilities in the near-infrared spectral region, particularly the $z$(1.0 $\mu$m), $J$ (1.2 $\mu$m), $H$(1.6 $\mu$m), and $K$ (2.1 $\mu$m) bands. On the other hand, precise measurements in the H$_2$O bands at 1.4 $\mu$m are possible with the {\it HST} WFC3 spectrograph. Similarly, molecular features of H$_2$O and other possible molecules (e.g. CH$_4$, CO, CO$2$), are also accessible with the warm {\it Spitzer} photometric bandpasses at 3.6 $\mu$m and 4.5$\mu$m. In the future, {\it JWST} will have the capability to revolutionize atmospheric characterization of hot H$_2$O-rich super-Earths with high-resolution spectroscopy over a much broader spectral range than is currently available. The prime super-Earths for characterization with JWST will be discovered in large numbers by upcoming surveys such as TESS (Ricker et al. 2014), CHEOPS (Broeg et al. 2013), and PLATO (Rauer et al. 2013) from space, and by several ground-based efforts (e.g. Snellen et al. 2012; Gillon et al. 2013a). 
   
 These new observational surveys would benefit from focusing on finding close-in transiting super-Earths orbiting bright stars. Very close-in super-Earths of SE2 and SE3 classes with T$_{\rm eq}$ $\geq 2000$ K would be particularly important, because the resulting planets could be successfully followed up with infrared telescopes to search for H$_2$O in their atmospheres as discussed above. In the present work, we simulated observations using the {\it HST} WFC3 spectrograph to identify parameters of super-Earth systems which would be most conducive for detecting their H$_2$O-rich atmospheres. We consider two scenarios of super-Earth host stars, a G dwarf and an M dwarf. We generally conclude that hotter planets around brighter or smaller host stars are more conducive for detecting H$_2$O-rich atmospheres. In principle, for M dwarf stellar hosts, super-Earths with T$_{\rm eq}$ as low as 500 K, such as GJ~1214b, could still be favorable for H$_2$O detection. However, the possibility of clouds in low temperature super-Earth atmospheres, especially for T$_{\rm eq}$ $\lesssim 1000$ K, complicate the interpretation of H$_2$O from spectra, as known from current observations of GJ~1214b. Consequently, very high temperature super-Earths (T$_{\rm eq}$ $\geq 2000$ K) orbiting bright stars currently present the best chances for constraining H$_2$O-rich atmospheres. Among currently known super-Earths, we find that the super-Earth 55 Cancri e (T$_{\rm eq} \sim 2000-2400$ K) presents the best chances for conclusively determining the presence of a H$_2$O-rich atmosphere in a super-Earth using spectroscopy in both transmission and thermal emission using {\it HST} WFC3.   
     
   Ultimately, detecting molecules like H$_2$O and other bio-signatures in super-Earth atmospheres might be possible to detect for a subset of super-Earths with {\it JWST}, but for earth-size planets the wait could be longer depending on the stellar hosts. In the meantime, however, a detailed survey of the abundances of H$_2$O in the most favorable sample of the hot super-Earth population is possible with existing facilities as discussed above.  The resulting constraints can help us estimate the likelihood of H$_2$O-rich atmospheres in habitable planets that are being discovered in parallel but whose atmospheres cannot be characterized in the near future. 
  \acknowledgements{NM acknowledges support from the Yale Center for Astronomy and Astrophysics (YCAA) at Yale University through the YCAA fellowship. SR acknowledges support by the National Science Foundation through Astronomy and Astrophysics Research Grant AST-1313268. We thank Peter McCullough, Avi Mandell, and Debra Fischer for helpful discussions. This research has made use of the Exoplanet Orbit Database and the Exoplanet Data Explorer at exoplanets.org.}
 \newline


\begin{thebibliography}{}
\bibitem[Abe(2011)]{Abe:11} Abe, Y., Abe-Ouchi, A., Sleep, N. H., \&  Zahnle, K. J. 2011, Astrobiology, 11, 443
\bibitem[Atreya(2010)]{Atreya:10} Atreya, S. K. 2010, Atmospheric Moons Galileo Would Have Loved. Galileo's Medicean Moons - Their Impact on 400 Years of Discovery (C. Barbieri et al., eds.),  Proceedings IAU Symposium No. 269, 2010. Cambridge, University Press.
\bibitem[Barclay(2013)]{Barclay:13} Barclay, T. et al. 2013, ApJ, 768, 101  
\bibitem[Batalha(2011)]{Batalha:11} Batalha, N. et al. 2011, ApJ, 729, 27 
\bibitem[Belu(2011)]{Belu:11} Belu, A. R. et al. 2011, A\&A, 525, A83
\bibitem[Belu(2013)]{Belu:13} Belu, A. R. et al. 2013, \apj, 768, 125
\bibitem[Bean(2011)]{Bean:11} Bean, J. et al. 2011, ApJ, 743, 92
\bibitem[Berta(2012)]{Berta:12} Berta, Z. et al. 2012, ApJ, 747, 35 
\bibitem[Broeg(2013)]{Broeg:13} Broeg, C. et al. 2013, Hot Planets and Cool Stars, Garching, Germany (Ed. Roberto Saglia), EPJ Web of Conferences, 47, 03005 (arXiv:1305.2270)	
\bibitem[Croll(2011)]{Croll:11} Croll, B. et al. 2011, ApJ, 736, 78 
\bibitem[Benneke(2012)]{Benneke:12} Bennekke, B. \& Seager, S. 2012, ApJ, 753, 100 
\bibitem[Benneke(2013)]{Benneke:13} Bennekke, B. \& Seager, S. 2013, arXiv1306.6325B
\bibitem[Borucki(2013)]{Borucki:13} Borucki, W. J. et al. 2013, Science, 340, 587 
\bibitem[Castan(2011)]{Castan:11} Castan, T. \& Menou, K. 2011, \apj, 743, L36
\bibitem[Charbonneau(2009)]{Charbonneau:09} Charbonneau, D. et al. 2009, Nature, 462, 891 
\bibitem[de Mooij(2012)]{de Mooij:12} de Mooij, E. J. W., et al. 2012, A\&A, 538, 46
\bibitem[Deming(2013)]{Deming:13} Deming, D., et al. 2013, \apj, 774, 95 
\bibitem[Demory(2011)]{Demory:11} Demory, B-O., et al. 2011, A\&A, 533A, 114
\bibitem[Demory(2012)]{Demory:12} Demory, B-O. et al. 2012, \apj, 751, 2, L28
\bibitem[Desert(2011)]{Desert:11} D\'esert, J.-M. et al. 2011, 197, 2011
\bibitem[Dragomir(2013)]{Dragomir:13} Dragomir, D. et al. 2013, \apj, 772, L2
\bibitem[Endl(2012)]{Endl:12} Endl, M. et al. 2012, \apj, 759, 19 
\bibitem[Ehrenreich(2012)]{Ehrenreich:12} Ehrenreich et al. 2012, A\&A 547, A18 
\bibitem[Fortney(2007)]{Fortney:07} Fortney, J. J., Marley, M. S., Barnes, J. W. 2007, \apj, 659, 1661
\bibitem[Fressin(2013)]{Fressin:13} Fressin, F. 2013, \apj, 766, 81
\bibitem[Gillon(2012)]{Gillon:12} Gillon, M. 2012, A\&A, 539A, 28
\bibitem[Gillon(2013a)]{Gillon:13a} Gillon, M., Jehin, E., Fumel, A., Magain, P., Queloz, D. 2013, Hot Planets and Cool Stars, Garching, Germany, Ed. R. Saglia, EPJ Web of Conferences, 47, id.03001
\bibitem[Gillon(2013b)]{Gillon:13b} Gillon, M. 2013b, arXiv: 1307.6722
\bibitem[Gong(2012)]{Gong:12} Gong, Y-X., \& Zhou, J-L. 2012, Res. in Astron. Astrophys., 12, 6, 678
\bibitem[Hedelt(2013)]{Hedelt:13} Hedelt, P. et al. 2013, A\&A, 553, A9
\bibitem[Heng(2012)]{Heng:12} Heng, K. \& Kopparla, P. 2012, \apj, 754, 60
\bibitem[Howe(2012)]{Howe:12} Howe, A. \& Burrows, A. 2012, \apj, 756, 176 
\bibitem[Howard(2012)]{Howard:12} Howard, A. et al. 2012, ApJS, 201, 15 
\bibitem[Johnson et al.(2012)]{Johnson:12} Johnson, T.~V., Mousis, O., Lunine, J.~I., \& Madhusudhan, N.\ 2012, \apj, 757, 192
\bibitem[Kaltenegger(2009)]{Kaltenegger:09} Kaltenegger, L. \& Traub, W. 2009, 698, 519
\bibitem[Kaltenegger(2013)]{Kaltenegger:13} Kaltenegger, L., Sasselov, D., Rugheimer, S. 2013, arXiv:1304.5058v1
\bibitem[Kasting(1993)]{Kasting:93} Kasting, J. F. 1993, 101, 108
\bibitem[Kipping(2013)]{Kipping:13} Kipping, D. M., Spiegel, D. S., \& Sasselov, D. D. 2013, MNRAS, 434, 1883
\bibitem[Kopparapu(2013a)]{Kopparapu:13a} Kopparapu, R. K. et al. 2013a, \apj, 765, 131 
\bibitem[Kopparapu(2013b)]{Kopparapu:13b} Kopparapu, R. K. et al. 2013b, \apj, 770, 82 
\bibitem[Kreidberg(2014)]{Kreidberg:14} Kreidberg, L. et al. 2014, Nature, 505, 69 
 \bibitem[Lecavelier(2008)]{Lecavelier:08} Lecavelier des Etangs, A., Pont, F., Vidal-Madjar1, A., \& Sing, D. 2008, A\&A, 481, L83
\bibitem[Lee(2012)]{Lee:12} Lee, J.-M., Fletcher, L. N., \& Irwin, P. G. J. 2012, MNRAS, 420, 170
\bibitem[Lepine(2011)]{Lepine:11} Lepine \& Gaidos 2011, AJ, 142, 138
\bibitem[Line(2012)]{Line:12} Line, M., et al. 2012, ApJ, 749, 93 
\bibitem[Leger(2009)]{Leger:09} Leger, A. et al. 2009, A\&A, 506, 287
\bibitem[Lodders(2002)]{Lodders:02} Lodders, K. 2002, \apj, 577, 974
\bibitem[Madhusudhan(2009)]{Madhusudhan:09} Madhusudhan, N. \& Seager, S. 2009, \apj, 707, 24
\bibitem[Madhusudhan(2011)]{Madhusudhan:11} Madhusudhan, N., et al. 2011, Nature, 469, 64
\bibitem[Madhusudhan(2011)]{Madhusudhan:11} Madhusudhan, N. \& Seager, S. 2011, \apj, 729, 41
\bibitem[Madhusudhan(2012)]{Madhusudhan:12} Madhusudhan, N. 2012, \apj, 758, 36
\bibitem[Madhusudhan(2012)]{Madhusudhan_et_al:12} Madhusudhan, N. et al. 2012, \apj, 759, L40
\bibitem[McCullough(2012)]{McCullough:12} McCullough, P. \& MacKenty, J. 2012, Instrument Science Report WFC3 2012-08, {\it HST}, Space Telescope Science Institute, STScI
\bibitem[Morley(2013)]{Morley:13} Morley, C. V. et al. 2013, \apj, 775, 33 
\bibitem[Moriarty(2014)]{Moriarty:14} Moriarty, J., Madhusudhan, N., \& Fischer, D. 2014, \apj, 787, 81 
\bibitem[Miller-Ricci(2009)]{Miller-Ricci:09} Miller-Ricci, E., Seager, S., \& Sasselov, D. 2009, \apj, 690, 1056
\bibitem[Miller-Ricci(2010)]{Miller-Ricci:10} Miller-Ricci, E. \& Fortney, J. J. 2010, \apj, 716, L74
\bibitem[Kempton(2012)]{Kempton:12} Kempton, E., Zahnle, K. \& Fortney, J. J. 2012, \apj, 745, 3 
\bibitem[Pickles(1998)]{Pickles:98} Pickles, A. J. 1998, PASP, 110, 863
\bibitem[Pont(2010)]{Pont:10} Pont, F., Knutson, H., Gilliland, R. L., Moutou, C. \& Charbonneau, D. 2008, \mnras, 385, 109 
\bibitem[Quintana(2014)]{Quintana:14} Quintana, E., et al. 2014, Science, 344, 277 
\bibitem[Rauer(2013)]{Rauer:13} Rauer, H. et al. 2013, Experimental Astronomy, submitted (arXiv:1310.0696)
\bibitem[Ricker(2014)]{Ricker:14} Ricker, G. et al. 2014, Proc. SPIE, Astronomical Telescopes + Instrumentation, submitted (arXiv:1406.0151)
\bibitem[Rogers(2010a)]{Rogers:10a} Rogers, L. A. \& Seager, S. 2010a, \apj, 712, 974 
\bibitem[Rogers(2010b)]{Rogers:10b} Rogers, L. A. \& Seager, S. 2010b, \apj, 716, 1208
\bibitem[Seager(2007)]{Seager:07} Seager, S., et al. 2007, \apj, 669, 1279 
\bibitem[Selsis(2007)]{Selsis:07} Selsis, F. 2007, Lectures in Astrobiology, Advances in Astrobiology and Biogeophysics, Springer-Verlag Berlin Heidelberg, 2007, p. 199
\bibitem[Snellen(2012)]{Snellen:12} Snellen, I., Stuik, R., Navarro, R., et al. 2012, Proc. SPIE, 8444, 84440I 
\bibitem[Snellen(2013)]{Snellen:13} Snellen, I. A. G. et al. 2013, \apj, 764, 182
\bibitem[Sotin(2007)]{Sotin:07} Sotin, C., Grasset, O., Mocquet, A. 2007, Icarus 191, 337.
\bibitem[Spiegel et al.(2009)]{Spiegel:09} Spiegel, D. S., Silverio, K., \& Burrows, A. 2009, \apj, 699, 1487
\bibitem[Sudarsky et al.(2003)]{Sudarsky:03} Sudarsky, D., Burrows, A., \& Hubeny, I. 2003, \apj, 588, 1121 
\bibitem[Valencia(2006)]{Valencia:06} Valencia, D., O'Connell, R. J., \& Sasselov, D. D. 2006, Icarus, 181, 545
\bibitem[Valencia(2010)]{Valencia:10} Valencia, D., Ikoma, M., Guillot, T., \& Nettelmann, N. 2010, A\&A, 516A, 20 
\bibitem[Valencia(2013)]{Valencia:13} Valencia, D., Guillot, T., Parmentier, V., \& Freedman, R. S. 2013, \apj, 775, 10  
\bibitem[von Braun(2011)]{von Braun:11} von Braun, K. et al., 2011, \apj, 740, 49  
\bibitem[Wagner(2012)]{Wagner:12} Wagner, F. W., Tosi, N., Sohl, F., Rauer, H., \& Spohn, T. 2012, A\&A, 541, 103 
\bibitem[Winn(2011)]{Winn:11} Winn, J. N., et al. 2011, \apj, 737, L18
\bibitem[Wright(2011)]{Wright:11} Wright, J. T. et al. 2011, PASP, 123, 412 
\end{thebibliography}
\end{document}